\documentclass[aps,prc,twocolumn,superscriptaddress,nofootinbib,longbibliography,floatfix,10pt]{revtex4-2}

\usepackage{graphicx}
\usepackage{dcolumn}
\usepackage{bm}
\usepackage{morefloats}
\usepackage{multirow}
\usepackage{amssymb}
\usepackage{amsmath}
\usepackage{xcolor}
\usepackage{longtable}
\usepackage{fix-cm}
\usepackage{mathptmx} 
\usepackage[T1]{fontenc}
\usepackage[colorlinks,allcolors=blue]{hyperref}

\makeatletter
\def\NAT@def@citea{\def\@citea{\NAT@separator}}
\makeatother

\begin{document}

\title{Study of multinucleon transfer mechanism in ${}^{160}\text{Gd}+{}^{186}\text{W}$ collisions in stochastic mean-field theory}

\author{S. Ayik}\email{ayik@tntech.edu}
\affiliation{Physics Department, Tennessee Technological University, Cookeville, TN 38505, USA}
\author{M. Arik}
\affiliation{Physics Department, Middle East Technical University, 06800 Ankara, Turkey}
\author{E. Erbayri}
\affiliation{Physics Department, Middle East Technical University, 06800 Ankara, Turkey}
\author{O. Yilmaz}
\affiliation{Physics Department, Middle East Technical University, 06800 Ankara, Turkey}
\author{A. S. Umar}
\affiliation{Department of Physics and Astronomy, Vanderbilt University, Nashville, TN 37235, USA}

\date{\today}

\begin{abstract}
Multinucleon transfer mechanism in collision of
${}^{160}\text{Gd}+{}^{186}\text{W}$ system is investigated in the framework of
quantal transport description, based on the stochastic mean-field (SMF) theory.
The SMF theory provides a microscopic approach for nuclear dynamics beyond the
time-dependent Hartree-Fock (TDHF) approach
by including mean-field fluctuations. Cross-sections for the primary fragment production
are determined in the quantal transport description and
compared with the available data.
\end{abstract}

\maketitle

\section{Introduction}

Production of heavy elements near the superheavy island with proton numbers
$Z>100$ embodies one of the greatest experimental and  theoretical
challenges~\cite{chatillon2006,herzberg2001,kozulin2012,kratz2013,watanabe2015,desai2019,adamian2010,adamian2010b,jiang2020,adamian2020,kalandarov2020,li2020b,adamian2021,itkis2022,heinz2022,wu2022}.
The most common approach for the production of these elements and their isotopes
is through fusion reactions. Historically, two distinct experimental approaches
have been employed to synthesize these nuclei, named based on their excitation
properties, as cold fusion reactions~\cite{hofmann2000} and hot fusion
reactions~\cite{oganessian2006,itkis2022}. The primary composite systems formed in these
reactions are at a relatively high excitation energy, which subsequently
de-excites by emitting neutrons, alpha particles, and secondary fission. This
results in an exceedingly small evaporation residue cross-sections that makes reaching to
heavier elements as well as the neutron rich isotopes of these elements very difficult. To
circumvent this obstacle multi-nucleon transfer reactions have been proposed as
an alternative. Such experiments recently focused on using actinide targets at near the
Coulomb barrier energies. Using this approach  it is possible to produce heavy primary
fragments at reasonably lower excitation energies. Consequently, such reactions
may provide a more efficient mechanism for production of heavy neutron-rich
isotopes, than fusion, fission, and fragmentation reactions.  In collision
involving deformed target nuclei multinucleon transfer depends on the collision
geometry.  In a typical collision, the system drifts toward symmetry. However,
for certain geometries the system may drift toward asymmetry, which is referred
to as inverse quasi-fission.   Multinucleon transfer mechanism has been studied
by employing a number of phenomenological approaches,  such as the
multidimensional Langevin
model~\cite{zagrebaev2006,zagrebaev2008,zagrebaev2008c,zagrebaev2011,karpov2017,saiko2019,saiko2022},
di-nuclear system model~\cite{feng2009,feng2009a,feng2017}, and the quantum molecular dynamics
approach~\cite{zhao2009,zhao2016,wang2016}.

To formulate a reliable description of the multiparticle transfer mechanism and
its dependence on the collision geometry it is essential to utilize microscopic
approaches.  Time-Dependent Hartree-Fock (TDHF) theory is a good candidate as
the basis for such a microscopic description to describe the evolution of
collective dynamics at low bombarding
energies~\cite{simenel2012,simenel2018,nakatsukasa2016,oberacker2014,umar2015a,umar2015c,umar2017,simenel2010,sekizawa2016,sekizawa2019}.
Despite its success, TDHF theory, based on the mean-field approach, only
describes the most probable path of the collision dynamics with small
fluctuations around it. By virtue of this limitation TDHF generally describes
the mean values of observables, such as the kinetic energy loss involving
one-body dissipation, but is unable to account for larger fluctuations and
dispersions of the fragment mass and charge distributions. In order to account
for these observables it is necessary to find a prescription that goes beyond
the mean-field approximation~\cite{tohyama2002a,tohyama2020,simenel2011,ayik2008,lacroix2014}.
One such approach is through the time-dependent
random phase approximation (TDRPA) developed by Balian and V\'en\'eroni, which
provides a consistent theory to computer larger fluctuations of the observables
going beyond mean-field. This method has been used to study multinucleon
transfer reactions in symmetric
systems~\cite{balian1985,balian1992,williams2018,godbey2020,godbey2020b}, it is inherently constrained to
compute the dispersion of charge and mass distributions in symmetric collisions.

The stochastic mean-field (SMF) theory, closely related to TDRPA, circumvents
this problem and facilitates further improvements to the  beyond mean field
approximation~\cite{ayik2008,lacroix2014}. The manuscript is organized as
follows: In Sec.~\ref{sec2}, we provide results of TDHF calculations for the
collisions of the ${}^{160}\text{Gd}+{}^{186}\text{W}$ system at
E$_{\text{c.m.}}=$502.6~MeV and  E$_{\text{c.m.}}=$461.9~MeV.
Section~\ref{sec3} briefly describes multi-nucleon transfer as constituted in
the quantal transport description of the SMF approach.  In this reaction both
projectile and target nuclei are two deformed isotopes between doubly closed
$^{132}$Sn and $^{208}$Pb shells (Z=50, N=82 and Z=82, N=126).  For most collision
geometries, initial mass asymmetry increases, which  cause reaction to be
characterized as inverse quasi-fission.  In Sec.~\ref{sec4}, we provide the
analysis of multinucleon transfer mechanism for the same reaction. The quantal
transport approach describes the production of primary isotopes and we compare
results with the available data~\cite{kozulin2017}.  The cross-section
distributions, mean values, and dispersions are determined without any
adjustable parameter employing a Skyrme energy density functional.  In
Sec.~\ref{sec5}, we summarize our results and provide conclusions.

\section{Mean-Field Properties}
\label{sec2}

The TDHF theory has been the primary microscopic tool for studying low-energy heavy-ion
reactions, including fusion, deep-inelastic collisions, and
quasifission\cite{simenel2012,nakatsukasa2016,oberacker2014,umar2015a,simenel2010,sekizawa2016,simenel2018,sekizawa2019}.
Since the theory is derived by the minimization of the time-dependent many-body action it
is deterministic in nature and provides the most probably reaction path for the system.
Namely, given a set of initial conditions for the reaction there is only one outcome
for the reaction. At some level distribution can be obtained by varying the initial
conditions (e.g. orbital angular momentum or the orientation angle for deformed nuclei).
However, these distributions are typically much narrower when compared with the experiment.
TDHF also provides a good description of one-body dissipation~\cite{washiyama2009,reinhard1988}.
In system ${}^{160}\text{Gd}+{}^{186}\text{W}$, initial ground states for the target and the projectile
have large prolate deformations.
Consequently, the reaction dynamics and the transfer of nucleons depend on the
relative alignment of the target and projectile. We refer to this as the
dependence on collision geometry.
For the reaction ${}^{160}\text{Gd}+{}^{186}\text{W}$, at initial energies
E$_{\text{c.m.}}$ = 461.9 MeV and E$_{\text{c.m.}}$ = 502.6 MeV, we explore
four distinct alignments of the target and the projectile.
Using the convention adopted in the work of Kedziora and Simenel, in Ref.~\cite{kedziora2010},
we denote initial orientation of either the projectile or the target principle deformation axis
to be in the beam direction with \textit{X},
and the case when their principle axis is perpendicular to the beam direction with
\textit{Y}.
As a result we are faced with four distinct orientation possibilities
for the target and the projectile, labeled as
\textit{YY, XX, XY, YX} corresponding to side-side, tip-tip, tip-side, and side-tip collision
geometries, respectively (please see Fig.~2 and Fig.~3~in Ref.~\cite{kedziora2010}.
Here, the first letter stands for the orientation of the lighter collision partner.
The calculations presented in the rest of the article employed the TDHF
code~\cite{umar1991a,umar2006c} using the SLy4d Skyrme energy density functional~\cite{kim1997},
with a box size of $60\times 60\times 36$~fm in the $x-y-z$ directions, respectively.
The results of our TDHF calculations for all of these
collision geometries are tabulated in Tables~\ref{tab1} and~\ref{tab2}
at two bombarding energies and for a range of initial orbital
angular momenta $\ell_i$. We denote the final values of mass and charge numbers
for the Gd-like fragments with
$A_1^f$, $Z_1^f$, and W-like fragment with $A_2^f$, $Z_2^f$, final total kinetic
energy lost (\textit{TKEL}), scattering angles in the c.m.
$\theta_{\text{c.m.}}$, and laboratory frame, $\theta_1^{lab}$ and $\theta_2^{lab}$.
These tables also include asymptotic values
of neutron $\sigma_{NN}$, proton $\sigma_{ZZ}$, mixed dispersions $\sigma_{NZ}$
and mass dispersions $\sigma_{AA}$, which will be discussed in Sec.~\ref{sec3}.
To economize on the computation time all quantities are evaluated in steps of
20 units of orbital angular momentum. The range of initial orbital angular momenta
is specified according to the angular position of detectors in the laboratory system.
The values of initial orbital angular momenta, which fall into detector acceptance
range of $25^\circ-65^\circ$ in the laboratory frame, are shown in
Table~\ref{tab1} and Table~\ref{tab2}. At the lower collision energy, E$_{\text{c.m.}}$ = 461.9~MeV,
only in the tip-tip and tip-side geometries produced fragments fall in the
acceptance range of detectors. As a result, in Table~\ref{tab2} only the tip-tip
and tip-side results are shown. Different collision geometries have qualitatively
distinct nucleon transfer mechanisms. These can be seen
more clearly by plotting the time evolution of neutron $N(t)$ and proton
$Z(t)$ numbers for Gd-like fragments or W-like fragments. In Fig.~\ref{fig1A} of
Appendix~\ref{appA}, time evolution of proton and neutron numbers of W-like
fragments in central collision of ${}^{160}\text{Gd}+{}^{186}\text{W}$ are
presented at different collision geometries.
\begin{table*}[!htb]
\caption{Results of the TDHF and SMF calculations for the
${}^{160}\text{Gd}+{}^{186}\text{W}$ system at $E_\text{c.m.}=502.6$~MeV in tip-tip
(XX), tip-side (XY), side-tip (YX) and side-side (YY) geometries.}
\label{tab1}
\begin{ruledtabular}
\begin{tabular}{c | c c c c c c c c c c c c c c r }
Geometry &$\ell_i$ $(\hbar)$ &$Z_1^f$ &$A_1^f$  &$Z_2^f$ & $A_2^f$ &$\ell_f$ $(\hbar)$
& $TKEL$    & $\sigma_{NN}$ & $\sigma_{ZZ}$ & $\sigma_{NZ}$ &$\sigma_{AA}$  & $\theta_\text{c.m.}$ & $\theta_{1}^{lab}$ & $\theta_{2}^{lab}$  \\
\hline

\multirow{7}{2.5em}{\textbf{XX}}
 &120 &61.1 &152.7 &76.9 &193.3 &117.1 &245.7 &5.8 &4.3 &3.9 &9.1 &120.8 &55.0 &24.1\\
 &140 &64.2 &160.5 &73.8 &185.5 &125.2 &237.0 &5.7 &4.3 &3.9 &9.0 &114.3 &51.1 &27.6\\
 &160 &65.8 &164.5 &72.2 &181.5 &134.0 &208.2 &5.6 &4.2 &3.7 &8.8 &111.8 &51.1 &30.0\\
 &180 &65.5 &164.0 &72.5 &182.0 &120.8 &173.5 &5.4 &4.1 &3.5 &8.4 &111.7 &53.4 &31.0\\
 &200 &65.6 &164.3 &72.4 &181.7 &138.0 &140.5 &5.1 &3.9 &3.2 &7.9 &107.5 &53.2 &33.9\\
 &220 &65.2 &163.5 &72.8 &182.5 &165.9 &114.7 &4.8 &3.5 &2.7 &7.1 &102.8 &52.2 &36.6\\
 &240 &64.5 &161.8 &73.5 &184.2 &202.3 &85.3 &4.1 &2.9 &1.9 &5.7 &98.7 &51.7 &39.1\\
\hline
\hline
\multirow{8}{2.5em}{\textbf{XY}}
 &120 &61.2 &152.6 &76.8 &193.4 &116.9 &212.3 &7.1 &4.9 &5.2 &11.3 &120.8 &57.9 &24.9\\
 &140 &64.7 &161.6 &73.3 &184.4 &132.8 &212.6 &6.9 &4.8 &5.0 &11.0 &113.6 &52.3 &28.8\\
 &160 &65.4 &163.3 &72.6 &182.7 &136.9 &186.0 &6.6 &4.5 &4.7 &10.4 &110.6 &52.3 &31.1\\
 &180 &65.3 &163.3 &72.7 &182.7 &141.2 &169.3 &6.2 &4.3 &4.3 &9.7 &106.8 &51.6 &33.2\\
 &200 &65.3 &163.3 &72.7 &182.7 &154.1 &155.3 &5.9 &4.1 &4.0 &9.1 &102.6 &50.3 &35.5\\
 &220 &65.1 &163.0 &72.9 &183.0 &170.6 &136.3 &5.4 &3.8 &3.5 &8.3 &99.0 &49.5 &37.6\\
 &240 &64.9 &162.3 &73.1 &183.7 &191.2 &113.8 &4.8 &3.4 &2.9 &7.1 &96.1 &49.0 &39.6\\
 &260 &64.4 &161.5 &73.6 &184.5 &218.8 &81.9 &3.9 &2.7 &1.9 &5.4 &94.1 &49.3 &41.4\\
\hline
\hline
\multirow{7}{2.5em}{\textbf{YX}}
 &120 &66.3 &166.3 &71.7 &179.6 &98.1 &177.3 &6.5 &4.6 &4.6 &10.3 &123.2 &57.5 &26.0\\
 &140 &66.1 &165.5 &71.9 &180.5 &117.4 &162.3 &6.2 &4.3 &4.2 &9.6 &117.2 &56.2 &29.0\\
 &160 &65.8 &164.7 &72.2 &181.3 &130.8 &146.4 &5.8 &4.1 &3.9 &9.0 &112.6 &55.3 &31.4\\
 &180 &65.9 &165.0 &72.1 &181.0 &148.2 &134.2 &5.5 &3.8 &3.5 &8.4 &107.9 &53.5 &33.9\\
 &200 &65.3 &163.8 &72.7 &182.2 &168.5 &122.0 &5.0 &3.5 &3.0 &7.4 &103.6 &52.2 &36.1\\
 &220 &64.5 &162.1 &73.5 &183.9 &186.0 &101.8 &4.4 &3.0 &2.3 &6.3 &100.4 &51.8 &37.9\\
 &240 &64.1 &161.0 &73.9 &185.0 &217.5 &73.5 &3.7 &2.5 &1.6 &4.9 &97.9 &51.8 &39.7\\
\hline
\hline
\multirow{6}{2.5em}{\textbf{YY}}
 &120 &64.1 &159.8 &73.9 &186.2 &106.7 &123.1 &5.4 &3.7 &3.5 &8.3 &125.7 &64.7 &25.4\\
 &140 &64.4 &160.8 &73.6 &185.2 &123.4 &113.8 &5.1 &3.5 &3.3 &7.7 &120.1 &62.0 &28.3\\
 &160 &64.6 &161.5 &73.4 &184.5 &141.2 &105.6 &4.8 &3.3 &2.9 &7.2 &115.0 &59.6 &30.9\\
 &180 &64.6 &161.6 &73.4 &184.4 &159.4 &95.6 &4.5 &3.1 &2.6 &6.5 &110.6 &57.7 &33.2\\
 &200 &64.4 &161.4 &73.6 &184.6 &175.4 &81.4 &4.0 &2.8 &2.1 &5.7 &106.9 &56.3 &35.3\\
 &220 &64.3 &161.2 &73.7 &184.8 &193.2 &66.3 &3.6 &2.4 &1.6 &4.8 &103.7 &55.3 &37.1\\
\end{tabular}
\end{ruledtabular}
\end{table*}

\begin{table*}[!htb]
\caption{Results of the TDHF and SMF calculations for the
${}^{160}\text{Gd}+{}^{186}\text{W}$ system at $E_\text{c.m.}=461.9$~MeV in tip-tip
(XX), tip-side (XY) geometries.}
\label{tab2}
\begin{ruledtabular}
\begin{tabular}{c | c c c c c c c c c c c c c c r }
Geometry &$\ell_i$ $(\hbar)$ &$Z_1^f$ &$A_1^f$  &$Z_2^f$ & $A_2^f$ &$\ell_f$ $(\hbar)$
& $TKEL$    & $\sigma_{NN}$ & $\sigma_{ZZ}$ & $\sigma_{NZ}$ &$\sigma_{AA}$  & $\theta_\text{c.m.}$ & $\theta_{1}^{lab}$ & $\theta_{2}^{lab}$  \\
\hline

\multirow{3}{2.5em}{\textbf{XX}}
 &140 &65.0 &162.6 &73.0 &183.4 &88.9 &97.7 &5.1 &3.8 &3.1 &7.7 &129.7 &65.8 &23.8\\
 &160 &64.9 &162.7 &73.1 &183.3 &116.1 &81.4 &4.7 &3.5 &2.7 &7.0 &123.3 &63.7 &27.1\\
 &180 &64.4 &161.5 &73.6 &184.5 &149.7 &56.7 &4.1 &3.0 &2.0 &5.8 &117.9 &62.6 &30.1\\
\hline
\hline
\multirow{2}{2.5em}{\textbf{XY}}
 &140 &64.6 &161.8 &73.4 &184.2 &117.2 &75.2 &4.8 &3.3 &2.8 &7.1 &126.3 &66.0 &25.7\\
 &160 &64.5 &161.7 &73.5 &184.3 &138.2 &59.2 &4.3 &3.0 &2.3 &6.1 &121.3 &64.3 &28.4\\
\end{tabular}
\end{ruledtabular}
\end{table*}

\begin{figure*}[!th]
\includegraphics*[width=15cm]{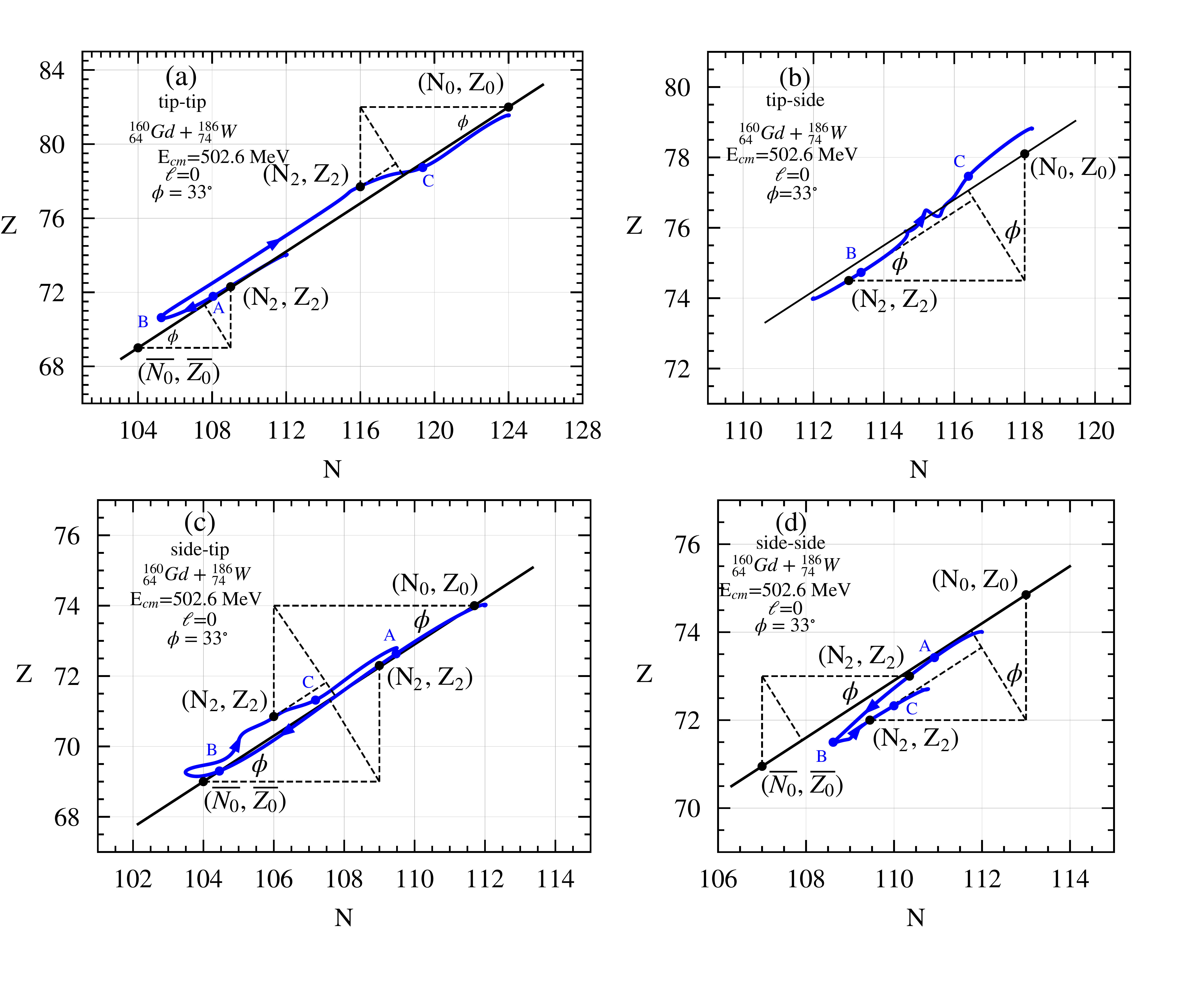}
\caption{Blue curves show drift path of W-like fragments in the head-on
collision of ${}^{160}\text{Gd}+{}^{186}\text{W}$ system at $E_\text{c.m.}=502.6$~MeV
in tip-tip (a), tip-side (b), side-tip (c) and side-side (d) geometries.
}
\label{fig1}
\end{figure*}

We can extract more useful information about nucleon transfer mechanisms by examining
drift paths. Drift paths are obtained by eliminating time from neutron $N(t)$
and proton numbers $Z(t)$ and plotting the result in $(N, Z)$ plane.
Drift paths in different collision geometries exhibit a different behavior
as a results of the shell effects on the dynamics, and contain specific information about the time
evolution of the mean values of macroscopic variables.
In Fig.~\ref{fig1} we plot drift paths for head-on collisions in different geometries
for W-like fragments (blue curves). In these figures, thick black lines denote set of
fragments with charge asymmetry value of $\frac{N-Z}{N+Z} = 0.20$. These lines are
called the iso-scalar path, which travel near the bottom of
stability valley. The iso-scalar path extends all the way from lead valley on
the upper end and toward the barium valley on the lower end, and it makes about
$\phi = 33^\circ$ with the horizontal neutron axis. We observe that in all
geometries, W-like fragments drift along iso-scalar direction with a charge
asymmetry of approximately 0.20. Fig.~\ref{fig1}(a) shows the drift path
in tip-tip (\textit{XX}) collisions. Initially, as seen usually in quasi-fission
reactions, W-like heavy fragments loses nucleons and the system drifts toward
symmetry. After this initial behavior, W-like heavy fragments turn back and by
gaining nucleons drift toward asymmetry. This kind of drift path is not very
usual and it is referred to as inverse quasi-fission reaction.
Fig.~\ref{fig1}(b) shows drift path in tip-side (\textit{XY}) collision. In this
case, drift path also indicates an inverse quasi-fission reaction.  In side-tip
(\textit{YX}) collision, shown in Fig.~\ref{fig1}(c), nucleon drift mechanism is
similar to the tip-tip (\textit{XX}) geometry. Initially, heavy fragment loses
neuron and protons and system drifts along iso-scalar path toward symmetry, subsequently
changing direction and drifting toward asymmetry. As seen in
Fig.~\ref{fig1}(d), in side-side (\textit{YY}) collision drift mechanism is analogous
to the one for side-tip collision.

\section{Quantal Diffusion of Nucleon Transfer}
\label{sec3}
\subsection{Langevin equations for multinucleon transfer}

In TDHF theory, with a prescribed set of initial conditions,
the many-body state is a single Slater determinant
and a unique single-particle density matrix, that is time-dependent
describing the deterministic reaction path for the dynamical system.
In beyond TDHF approaches, introducing additional correlations are
typically represented by a superposition of Slater determinants.
In the SMF theory, correlations are introduced as fluctuations of the initial
state, which constitute an ensemble of single-particle density matrices~\cite{ayik2008,lacroix2014}.
For each of these single-particle density matrices in the ensemble, the time-evolution
reduces to the TDHF equations initialized by the self-consistent Hamiltonian of the particular event.
In constructing the fluctuations of these initial density matrices SMF
employs a Gaussian distribution of the random elements with
variances that are specified with the requirement that
ensemble average of the one-body operator dispersions of the initial state
are the same as the ones obtained in the quantal expressions in the mean-field approach.

For low-energy heavy-ion collisions at energies near the Coulomb barrier the dynamical
system generally maintains a di-nuclear configuration. In these cases instead of
generating an ensemble of mean-field events one can formulate a more straight forward
transport approach. Using the window dynamics of the di-nuclear system one can
do a geometric projection of the SMF approach to obtain
Langevin equations for the relevant macroscopic
variables. For the derivations of the quantal diffusion formalism
and the utilization of window dynamics we refer the reader to earlier
references~\cite{ayik2017,ayik2018,yilmaz2018,ayik2019,sekizawa2020,ayik2020b,yilmaz2020}.
Neutron and proton number of the projectile-like or target-like fragment
are chosen as the relevant macroscopic variables to formulate the
diffusion formalism.
For the system at hand, we take neutron $N_2^{\lambda } (t)$ and
proton $Z_2^{\lambda } (t)$ numbers of the W-like fragments as relevant
macroscopic variables.
For each event, $\lambda$, the fragment neutron and proton numbers are
obtained by integrating the density on the left and right of the window.
During the contact phase, fragment proton and neutron numbers fluctuate
between events as a result of random nucleon flux
through the window. These numbers can be decomposed as fluctuations about the
mean values as
$N_2^{\lambda}(t)=N_2(t)+\mathit{\delta N}_1^{\lambda }(t)$ and $Z_2^{\lambda
}(t)=Z_2(t)+\mathit{\delta Z}_2^{\lambda }(t)$. Here, $N_2(t)$ and $Z_2(t)$ are
the mean values obtained over the ensemble of SMF events.

These mean values can be deduced from the mean-field TDHF calculations for small
amplitude fluctuations. In the  quantal diffusion approach,
for small amplitude
fluctuations,  neutron $\mathit{\delta N}_2^{\lambda }(t)$ and proton
$\mathit{\delta Z}_2^{\lambda }(t)$ numbers evolve according to a coupled linear
set of quantal Langevin equations,
\begin{align}\label{eq1}
\frac{d}{dt} \left(\begin{array}{c} {\delta Z_{2}^{\lambda } (t)} \\ {\delta N_{2}^{\lambda } (t)}
\end{array}\right)=&\left(\begin{array}{c} {\frac{\partial v_{p} }{\partial
Z_{2} } \left(Z_{2}^{\lambda } -Z_{2} \right)+\frac{\partial v_{p} }{\partial
N_{2} } \left(N_{2}^{\lambda } -N_{2} \right)} \\ {\frac{\partial v_{n}
}{\partial Z_{2} } \left(Z_{2}^{\lambda } -Z_{2} \right)+\frac{\partial v_{n}
}{\partial N_{2} } \left(N_{2}^{\lambda } -N_{2} \right)}
\end{array}\right)\nonumber\\ &+\left(\begin{array}{c} {\delta v_{p}^{\lambda }
(t)} \\ {\delta v_{n}^{\lambda } (t)} \end{array}\right).
\end{align}
Quantities $v_{\alpha}^{\lambda}(t)=v_{\alpha}(t)+\delta
v_{\alpha}^{\lambda }(t)$ denote the drift coefficients for protons and neutrons,
with the mean values and the fluctuating parts given by $v_{\alpha }(t)$ and
$\mathit{\delta v}_{\alpha }^{\lambda }(t)$, respectively. Here, the index $\alpha$
labels protons and neutrons. Drift coefficients
$v_{\alpha}^{\lambda}(t)$, denote the rate of flux for protons and neutrons through
the window in event $\lambda $. The linear limit of Langevin description
employed here is a good approximation when the driving potential energy
is more or less harmonic around equilibrium values of the mass and charge asymmetry.
The rate of change of neutron and proton numbers determines the
mean values of drift coefficients. For W-like fragments these are shown in Fig.~\ref{fig1A} of the
Appendix~\ref{appA}. Quantal expressions for the stochastic parts of
the drift coefficients $\mathit{\delta v}_{\alpha }^{\lambda }(t)$ can be found in
Ref.~\cite{ayik2020,ayik2023}.

\subsection{Quantal Diffusion Coefficients}

Stochastic parts of the drift coefficients $\delta v^{\lambda}_{\text{p}} (t)$
and $\delta v^{\lambda}_{\text{n}}(t)$ are the primary generator of
fluctuations in charge and mass asymmetry degrees of freedom. In the
SMF theory, these stochastic parts of drift coefficients are Gaussian random
distributions centered about a zero mean value, $\delta \overline{v}_{p,n}^{\lambda }(t)=0$.
The auto-correlation functions of the
stochastic drift coefficient, written as an integration over the evolution history, give the
diffusion coefficients $D_{\alpha\alpha} (t)$ for proton and neutron transfer
\begin{align}\label{eq2}
\int_{0}^{t}dt'\overline{\delta v_{\alpha }^{\lambda } (t')\delta v_{\alpha
}^{\lambda}(t)} =D_{\alpha \alpha }(t)\;.
\end{align}
\begin{figure*}[!htb]
\includegraphics*[width=15cm]{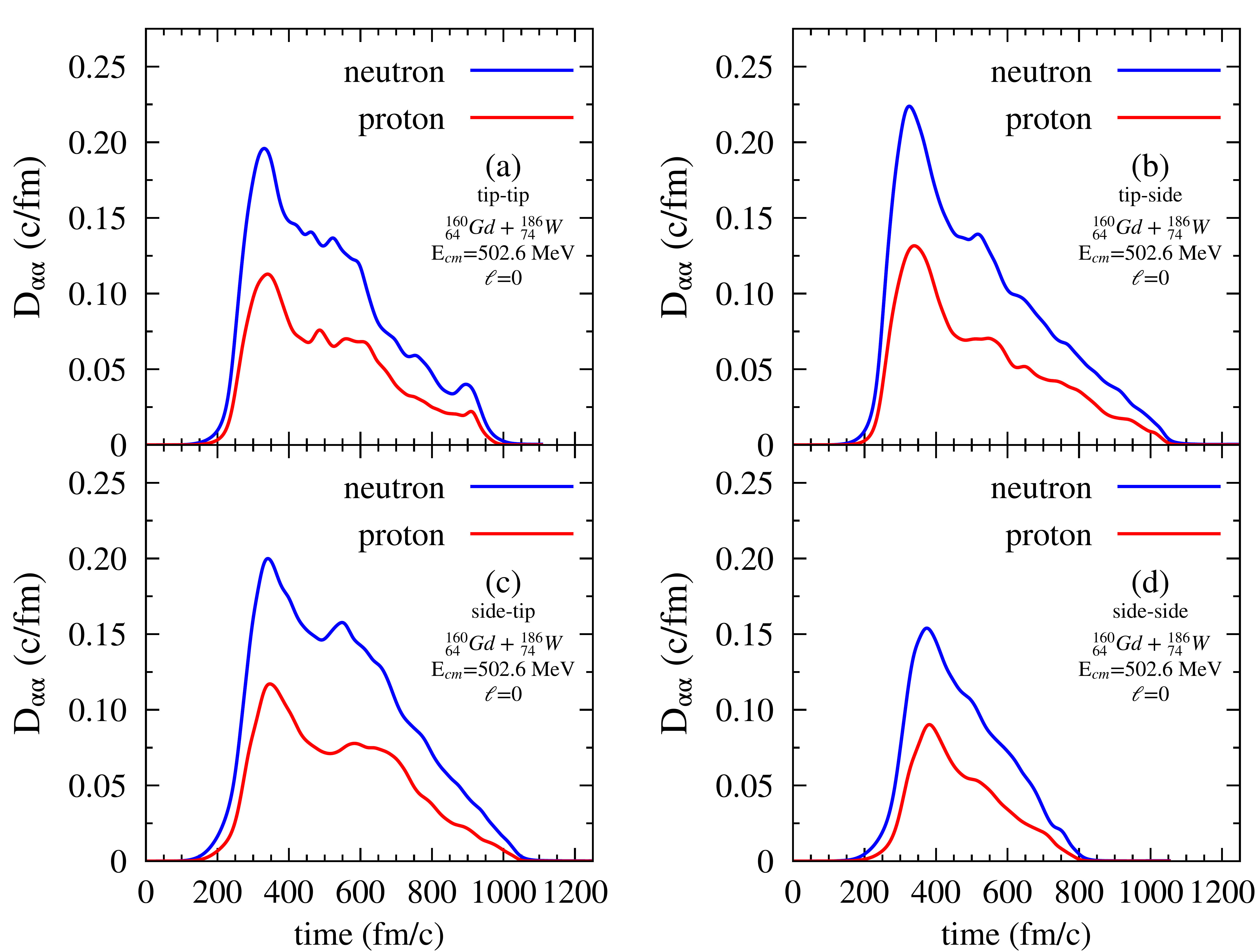}
\caption{Neutron and proton diffusion coefficient in the head-on collision of
${}^{160}\text{Gd}+{}^{186}\text{W}$ system at $E_\text{c.m.}=502.6$~MeV  in tip-tip
(a), tip-side (b), side-tip (c) and side-side (d) collision geometries.}
\label{fig2}
\end{figure*}

In the most general formulation, the diffusion coefficients are over a complete set of particle-hole states.
In the diabatic limit we can get rid of the particle states by utilizing closure relations.
As a consequence the diffusion coefficients are then obtained solely in terms of the occupied
single-particle states of the TDHF calculations, which provides a significant simplification.
Our previous publications provide all the explicit expressions for the diffusion
coefficients~\cite{ayik2017,ayik2018,yilmaz2018,ayik2019,sekizawa2020,ayik2020b,yilmaz2020,ayik2020,ayik2023}.
The analysis of these coefficients are also provided in Appendix~B in Ref.~\cite{ayik2017}.
The determination of the diffusion coefficients by virtue of the mean-field properties is
consistent with the fluctuation dissipation theorem of non-equilibrium
statistical mechanics and it significantly uncomplicates the calculation of quantal
diffusion coefficients. Quantal properties such as shell effects, Pauli principle,
and the effect of the unrestricted collision geometry, are included in these
diffusion coefficients without any adjustable parameters.
We point out that there is a close analogy
between the quantal expression and the classical diffusion coefficient for a
random walk problem~\cite{gardiner1991,weiss1999,risken1996}.
The direct part is given as the sum of the nucleon
currents across the window from the projectile-like fragment to the target-like fragment
and from the target-like fragment to the projectile-like fragment, integrated over the memory. This is analogous to the random walk
problem, in which the diffusion coefficient is given by the sum of the rate of
the forward and backward steps.
Pauli blocking effects in the transfer mechanism does not have a classical
counterpart and is represented in the second part of the quantal diffusion equations.
This is illustrated in Fig.~\ref{fig2}, where we plot
the neutron and proton diffusion coefficients in head-on collisions of
${}^{160}\text{Gd}+{}^{186}\text{W}$ system at $E_{\text{c.m.}}=502.6$~MeV for
different collision geometries.

\begin{figure*}[!htb]
\includegraphics*[width=15cm]{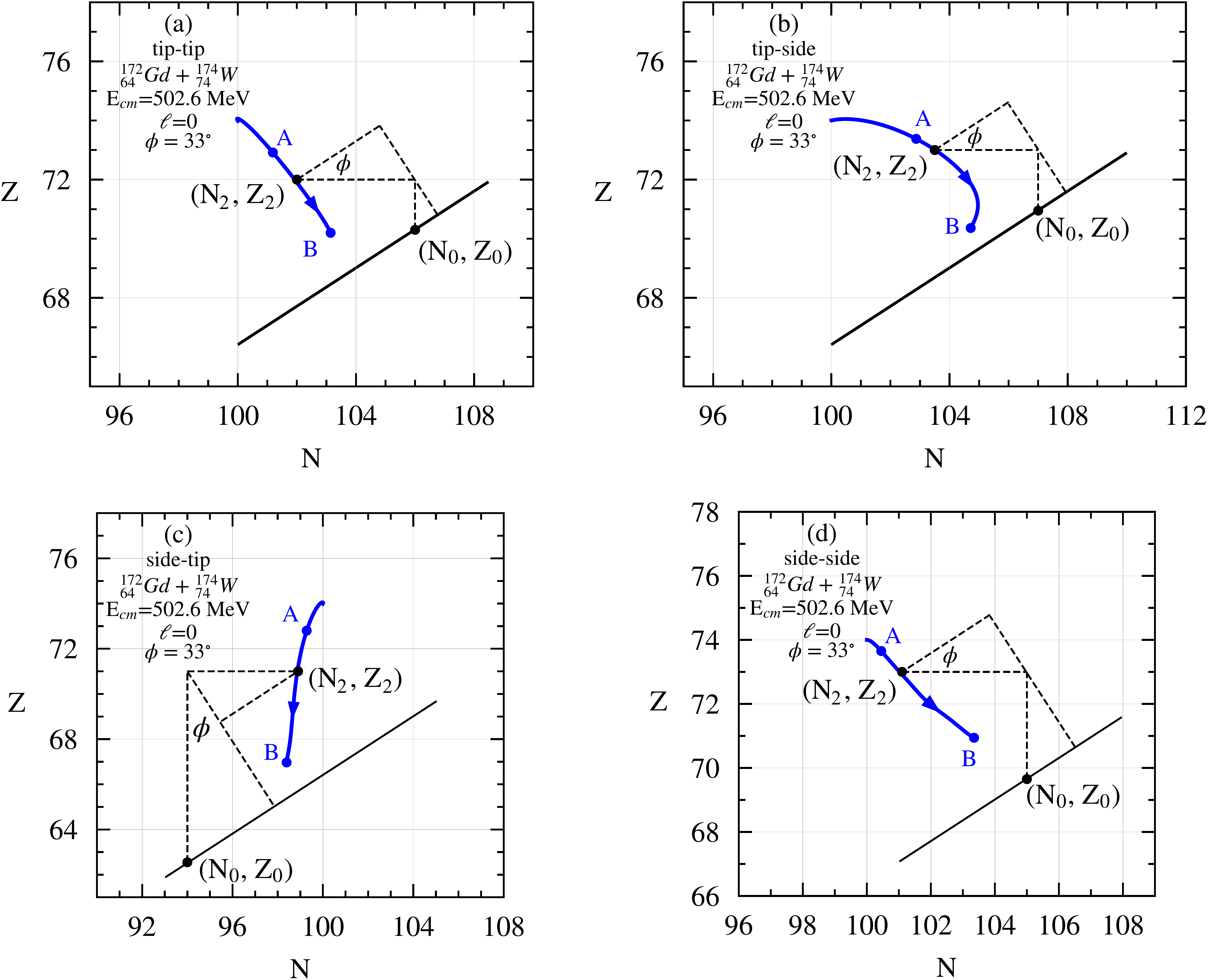}
\caption{Blue curves show drift path of W-like fragments in the head-on
collision of ${}^{172}\text{Gd}+{}^{174}\text{W}$ system at $E_\text{c.m.}=502.6$~MeV
in tip-tip (a), tip-side (b), side-tip (c) and side-side (d) geometries.
}
\label{fig3}
\end{figure*}

\begin{figure*}[!htb]
\includegraphics*[width=14cm]{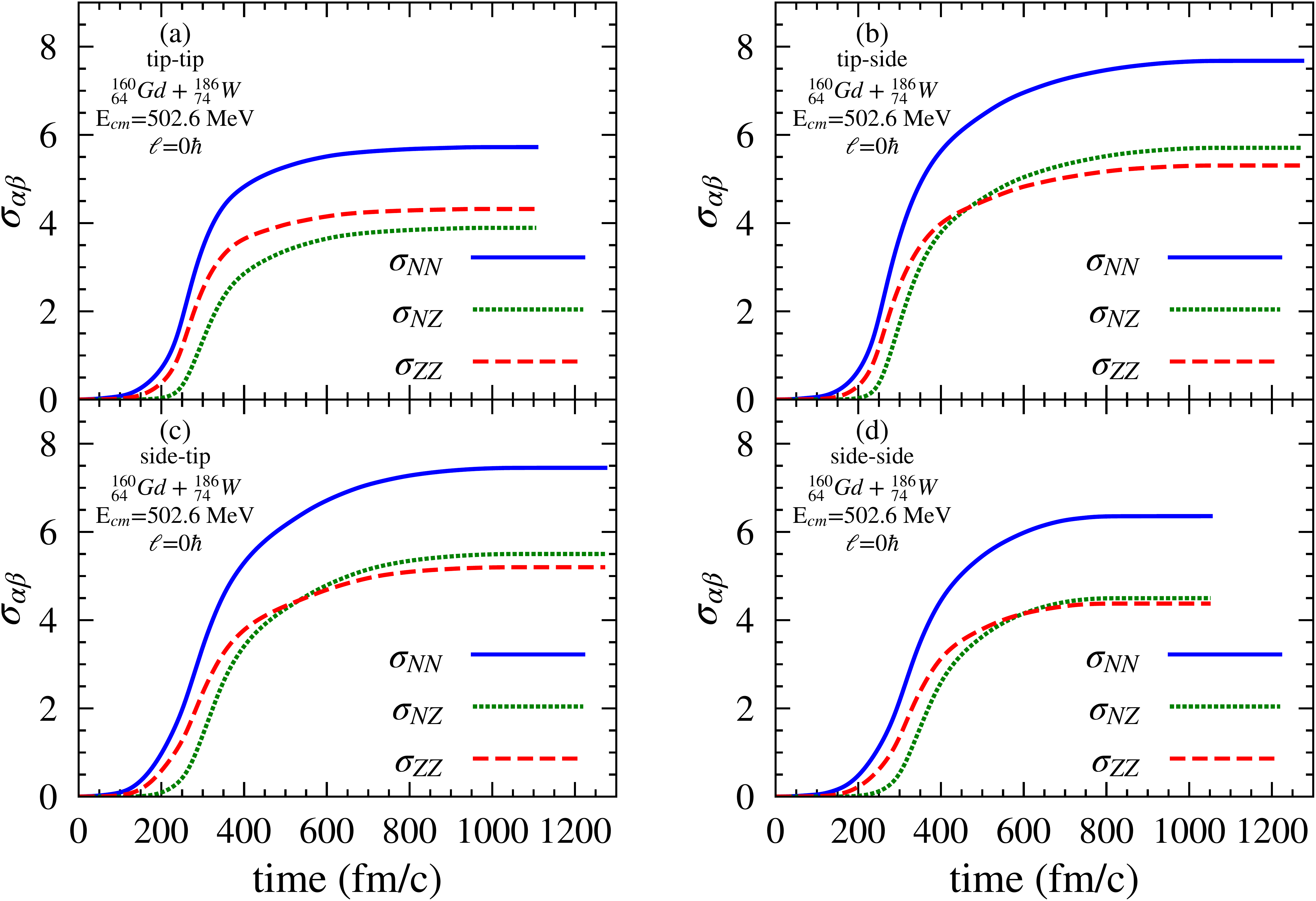}
\caption{Neutron, proton and mixed variances as a function of time in the head-on collision of
${}^{160} \text{Gd}+{}^{186} \text{W}$ system at $E_\text{c.m.}=502.6$~MeV  in tip-tip
(a), tip-side (b), side-tip (c) and side-side (d) geometries.}
\label{fig4}
\end{figure*}

\subsection{Potential energy of di-nuclear system}
Nucleon drift mechanism as well as dispersions of fragment distributions are
determined in terms of two competing effects: (i) Nucleon diffusion tends to
increase dispersion of distribution functions and (ii) Potential energy of
di-nuclear system $U(N_2,Z_2)$ on neutron and proton plane that controls the
mean nucleon drift and determines saturation values of dispersions. Potential
energy of the di-nuclear system consists mainly of  electrostatic
energy, symmetry energy, surface energy, and centrifugal energy.
The TDHF theory encompasses
different energy contributions at the microscopic level. Furthermore, TDHF
calculations illustrate potential energy depends on the geometry of di-nuclear
system. It is possible to extract useful information about potential energy with
the help of Einstein relation in the over damped
limit~\cite{ayik2017,ayik2018,yilmaz2018,ayik2019,sekizawa2020}. In the over
damped limit, drift coefficients are related to the potential energy surface in
$(N,Z)$ plane as
\begin{subequations}
\label{eq3}
\begin{align}
 v_{n}(t)  &=-\frac{D_{NN}(t) }{T^{*} } \frac{\partial }{\partial N_{2} } U(N_2,Z_2)\;,\label{eq3a}\\
 v_{z}(t) &=-\frac{D_{ZZ}(t) }{T^{*} } \frac{\partial }{\partial Z_{2}} U(N_2,Z_2)\;,\label{eq3b}
\end{align}
\end{subequations}
where $T^{\ast }$ stands for the effective temperature of the system, and
$N_2,Z_2$ indicate neutron-proton numbers of W-like fragments in the di-nuclear
system. Drift information of ${}^{160}\text{Gd}+{}^{186}\text{W}$ system give
information about potential energy only in the iso-scalar direction.  To obtain
information about potential energy in direction perpendicular to the stability
valley, we consider collision of a neighboring system
${}^{172}\text{Gd}+{}^{174}\text{W}$ at $E_\text{c.m.}=502.6$~MeV.
Fig.~\ref{fig3} shows drift path of W-like fragments in head-on collisions in
different geometries. In all geometries, system rapidly follows a path nearly
perpendicular direction to the stability valley to reach charge asymmetry
equilibrium.  We refer the direction perpendicular to stability valley as the
iso-vector direction. Evolution in iso-vector direction is accompanied by drift
along iso-scalar path toward symmetry or asymmetry. At the end of relatively
short contact time, system separates before reaching the local equilibrium
state.

In collisions of ${}^{160}\text{Gd}+{}^{186}\text{W}$ system, heavier local
equilibrium state is in vicinity of lead valley with neutron and proton numbers
around $N_0 = 124$, $Z_0 = 82$, while the lighter local equilibrium state is in the
neighborhood of barium valley having neutron and proton numbers near $\overline{N_0}
= 84$, $\overline{Z_0} = 56$. These nuclei are located on the iso-scalar path
with the charge asymmetry 0.20. The drift information of these two
similar systems, when combined, can provide an approximate description of potential
energy surface of the di-nuclear system relative to the equilibrium value of
potential energy in terms of two parabolic forms,
\begin{align}\label{eq4}
U(N_{2} ,Z_{2} )=\frac{1}{2} aR_{S}^{2} (N_{2} ,Z_{2} )+\frac{1}{2} bR_{V}^{2} (N_{2} ,Z_{2} )\;.
\end{align}
Here, $R_S(N_2,Z_2)$ and  $R_V(N_2,Z_2)$ represent perpendicular distances of a
fragment with neutron and proton numbers $(N_2,Z_2)$ from the isoscalar path
and from the local equilibrium state along the isoscalar path, respectively.

As a consequence of the sharp increase in asymmetry energy, we anticipate the isovector
curvature parameter to be considerably greater than the corresponding
isoscalar curvature parameter.
From the geometry of Fig.~\ref{fig1} and Fig.~\ref{fig3}, we express iso-vector
and iso-scalar distances in terms of neutron and proton number of the fragment
and neutron and proton numbers of equilibrium states. When drift is in asymmetry
direction iso-scalar distance is given by $R_{V}(t)= \left[~ N_0-N_2(t) ~\right]
\text{cos}\phi + \left[~ Z_0-Z_2(t) ~\right] \text{sin}\phi$, and for drift in
symmetry direction it is given by $\overline{R_{V}}(t)= \left[~ N_2(t) -
\overline{N_0} ~\right] \text{cos}\phi + \left[~ Z_2 (t) - \overline{Z_0} ~
\right] \text{sin}\phi$. In both cases iso-vector distance is given by
$R_{S}(t)= \left[~  N_0-N_2(t) ~\right] \text{sin}\phi -  \left[~ Z_0-Z_2(t)~
\right] \text{cos}\phi$. The angle $\phi$ is the angle between the iso-scalar
path and $N$-axis which is about $\phi = 33^{\circ}$.

Mean drift coefficients and diffusion coefficients are determined from the TDHF
and SMF calculations, using Einstein relations in Eq.~(\ref{eq3}), we can
determine the reduced curvature coefficients $\alpha = a/T^{\ast}$ and $\beta =
b/T^{\ast}$. Only ratios of the curvature parameters and the effective
temperature appear. As a result, the effective temperature is not a parameter
in the description. The reduced curvature parameters in each collision geometry
vary in time due the shell structure of TDHF description. In calculations of
dispersion values we employ constant curvature parameters, which  are determined
by averaging over suitable time intervals when the overlap between the colliding nuclei
are sufficiently large. When drift occurs toward asymmetry, the averaged
value of the iso-scalar reduced curvature parameter over a time interval $t_B$
and  $t_C$ is determined as,

\begin{align}\label{eq5}
\beta^{asym} = \frac {1} {R_{V}^{asym}} \int _{t_B}^{t_C}\left(\frac{v_{n} (t)\cos
\phi }{D_{NN} (t)} +\frac{v_{p} (t)\sin \phi }{D_{ZZ} (t)} \right)  dt\;,
\end{align}
where the integrated value of isoscalar distance for drift towards asymmetry is given by
\begin{align}\label{eq6}
R_V^{asym}=\int _{t_B}^{t_C}\left\{ \left[~ N_0 - N_2(t)~ \right] \text{cos}\phi + \left[~ Z_0 - Z_2(t) ~\right]  \text{sin}\phi
\right\} dt\;.
\end{align}

When drift occurs toward symmetry, we can determine the averaged value of the
iso-scalar reduced curvature parameter over a time interval $t_A$ and $t_B$ as,

\begin{align}\label{eq7}
\beta^{sym} = - \frac {1} {R_{V}^{sym}} \int _{t_A}^{t_B}\left(\frac{v_{n} (t)\cos
\phi }{D_{NN} (t)} +\frac{v_{p} (t)\sin \phi }{D_{ZZ} (t)} \right)  dt\;,
\end{align}
where the integrated value of isoscalar distance for drift towards symmetry is given by
\begin{align}\label{eq8}
R_V^{sym}=\int _{t_A}^{t_B}\left\{ \left[~ N_2(t) - \overline{N_0} ~\right] \text{cos}\phi + \left[~ Z_2(t) - \overline{Z_0} ~\right] \text{sin}\phi
\right\} dt\;.
\end{align}

\begin{table}[!htb]
\caption{Calculation of $\beta$ curvature parameter using ${}^{160}\text{Gd}+{}^{186}\text{W}$ system at
$E_\text{c.m.}=502.6$~MeV in tip-tip(XX), tip-side (XY), side-tip (YX) and side-side (YY) geometries.
The time intervals $t_A$, $t_B$ and $t_C$ required to calculate isoscalar curvature parameters are shown in Fig.~\ref{fig1A}.}
\label{tab3}
\begin{ruledtabular}
\begin{tabular}{ c | c | c | c | c | c | c }
${}^{160}\text{Gd}+{}^{186}\text{W}$ &$t_A$ &$t_B$ &$t_C$ &$\beta^{sym}_{AB}$
&$\beta^{asym}_{BC}$ &$\overline{\beta}= (\beta^{sym}_{AB} + \beta^{asym}_{BC})/2 $ \\
\hline
tip-tip &260 &250 &800 &0.010 &0.034 &0.022 \\
\hline
tip-side &--- &250 &900 &--- & 0.008 &0.008 \\
\hline
side-tip &260 &380 &900 &0.016 &0.004 &0.010 \\
\hline
side-side &300 &450 &750 &0.007 &0.010 &0.009
\end{tabular}
\end{ruledtabular}
\end{table}
These expressions can be used to calculate the averaged values of isoscalar reduced
curvature parameters in different geometries. In Table~\ref{tab3}, the
calculated values of isoscalar reduced curvature parameters for different
geometries are given. In tip-tip, side-tip and side-side geometries, initially
drift towards symmetry is observed, followed by a drift towards asymmetry. For
these geometries, we determine the isoscalar curvature parameter by taking the
average of drift towards symmetry and drift towards asymmetry part, given as
$\overline{\beta}= (\beta^{sym}_{AB} + \beta^{asym}_{BC})/2 $.

\begin{table}[!htb]
\caption{Calculation of $\alpha$ curvature parameter using ${}^{172}\text{Gd}+{}^{174}\text{W}$ system at $E_\text{c.m.}=502.6$~MeV in tip-tip (XX),
 tip-side (XY), side-tip (YX) and side-side (YY) geometries.
The time intervals $t_A$ and $t_B$ required to calculate isovector curvature parameters are shown in Fig.~\ref{fig2A}.}
\label{tab4}
\begin{ruledtabular}
\begin{tabular}{ c | c | c | c }
${}^{172}\text{Gd}+{}^{174}\text{W}$ &$t_A$ &$t_B$ &$\alpha$ \\
\hline
tip-tip &180 &260 &0.113 \\
\hline
tip-side &250 &400 &0.133 \\
\hline
side-tip &200 &350 &0.127 \\
\hline
side-side &250 &450 &0.143 \\
\end{tabular}
\end{ruledtabular}
\end{table}

We estimate the iso-vector reduced curvature parameters in different geometries
from the drift paths of ${}^{172}\text{Gd}+{}^{174}\text{W}$ system at
$E_\text{c.m.}=502.6$~MeV by averaging over time interval $t_A$ and $t_B$ as,

\begin{align}\label{eq9}
\alpha =  \frac {1} {R_{S}} \int _{t_A}^{t_B}\left(\frac{v_{n} (t)\sin
\phi }{D_{NN} (t)} - \frac{v_{p} (t)\cos \phi }{D_{ZZ} (t)} \right)  dt\;,
\end{align}
where the integrated value of iso-vector distance is given by
\begin{align}\label{eq10}
R_S=\int _{t_A}^{t_B}\left\{ \left[ ~N_0 - N_2(t) ~\right] \text{sin}\phi - \left[~Z_0 - Z_2(t)~\right] \text{cos}\phi
\right\} dt\;.
\end{align}
In Eq.~(\ref{eq9}), diffusion coefficients for ${}^{172}\text{Gd}+{}^{174}\text{W}$ system at $E_\text{c.m.}=502.6$~MeV are plotted in Fig.~\ref{fig3A}.

We can use these expressions in calculating averaged values of isovector reduced
curvature parameters in different geometries. In Table~\ref{tab4}, the
calculated values of isovector reduced curvature parameters for different
geometries are given.

Potential energy surface in (N-Z) plane should not depend on the centrifugal
potential energy and excitation energy of the di-nuclear system. Therefore, in
${}^{160}\text{Gd}+{}^{186}\text{W}$ system at $E_\text{c.m.}=461.9$~MeV, we
employ the same curvature parameters which are determined at
$E_\text{c.m.}=502.6$~MeV. Since the drift coefficients have analytical form, we
can immediately determine their derivatives to find,

\begin{align}
\frac{\partial \nu _{n}}{\partial N_{2} } &=-D_{NN} \left(\alpha \sin ^{2} \phi +\beta \cos ^{2} \phi \right)\label{eq11}\;,\\
\frac{\partial \nu _{z} }{\partial Z_{2} } &=-D_{ZZ} \left(\alpha \cos ^{2} \phi +\beta \sin ^{2} \phi \right)\label{eq12}\;,\\
\frac{\partial \nu _{n} }{\partial Z_{2} } &=-D_{NN} \left(\beta -\alpha \right)\sin \phi \cos \phi\label{eq13}\;,\\
\frac{\partial \nu _{z} }{\partial N_{2} } &=-D_{ZZ} \left(\beta -\alpha \right)\sin \phi \cos \phi\label{eq14}\;.
\end{align}

The curvature parameter $\alpha$ perpendicular to the beta stability valley is
much larger than the curvature parameter $\beta$ along the stability valley.
Consequently, $\beta$  does not have an appreciable effect on the derivatives of
the drift coefficients.  These derivative expressions are needed to calculate
neutron, proton, and mixed dispersions, as discussed in the following section.

\subsection{Fragment Probability Distributions}

In general, the combined probability distribution function $P_\ell (N,Z)$ for producing
a fragment with neutron $N$ and proton $Z$ is obtained by producing a
large number of solutions of Langevin Eq.~(\ref{eq1}). The equivalence between the
Langevin Equation and the Fokker-Planck equation, for the
distribution function of the macroscopic variables~\cite{weiss1999}, is
of common knowledge. When the drift coefficients are linear functions of macroscopic variables, as
in the case of Eq.~(\ref{eq1}), the proton and neutron distribution function for
initial angular momentum $\ell$ is given as a correlated Gaussian function
described by the mean values, and the neutron, proton, and mixed dispersion, as

\begin{align}\label{eq15}
P_{\ell} (N,Z)=\frac{1}{2\pi \sigma _{NN}(\ell)\sigma _{ZZ}(\ell)\sqrt{1-\rho
_{\ell}^{2}}}\exp\left(-C_{\ell}\right)\;.
\end{align}
Here, the exponent $C_{\ell}$ is given by
\begin{align}\label{eq16}
C_{\ell} =\frac{1}{2\left(1-\rho_{\ell}^{2} \right)}
&\left[\left(\frac{Z-Z_{\ell} }{\sigma _{ZZ} (\ell)} \right)^{2} -2\rho_\ell
\left(\frac{Z-Z_{\ell} }{\sigma _{ZZ} (\ell)} \right)\left(\frac{N-N_{\ell}
}{\sigma _{NN} (\ell)} \right)\right.\nonumber\\ &\left.+\left(\frac{N-N_{\ell}
}{\sigma _{NN} (\ell)} \right)^{2} \right]\;,
\end{align}
with the correlation coefficient defined as $\rho _\ell=\sigma
_{\text{NZ}}^2(\ell)/(\sigma _\text{ZZ}(\ell)\sigma _{\text{NN}}(\ell))$.
Quantities $N_\ell=\overline{N}_\ell^{\lambda }$,
$Z_\ell=\overline{Z}_\ell^{\lambda }$ denote the mean neutron and proton numbers
of the target-like or project-like fragments. These mean values are obtained
by performing TDHF calculations.
It is possible to
deduce coupled differential equations for variances $\sigma
_{\text{NN}}^2(\ell)=\overline{\mathit{\delta N}^{\lambda}\mathit{\delta
N}^{\lambda }}$,  $\sigma _{\text{ZZ}}^2(\ell)=\overline{\mathit{\delta Z}^{\lambda
}\mathit{\delta Z}^{\lambda }}$, and co-variances  $\sigma
_{\text{NZ}}^2(\ell)=\overline{\mathit{\delta N}^{\lambda }\mathit{\delta
Z}^{\lambda }}$ by multiplying Langevin Eq.~(\ref{eq1}) with $\mathit{\delta
N}^{\lambda }$, ${\delta Z}^{\lambda }$ and carrying out the average over the
ensemble generated from the solution of the Langevin equation. These coupled
equations were obtained in
Refs.~\cite{ayik2017,ayik2018,yilmaz2018,ayik2019,ayik2019b,sekizawa2020,ayik2020b}.
We provide these
differential equations here again for completeness~\cite{schroder1981},
\begin{align}\label{eq17}
\frac{\partial}{\partial t } {\sigma}^2_{NN} = 2 \frac{\partial
\nu_{n}}{\partial N_2} \sigma^2_{NN} + 2 \frac{\partial \nu_{n}}{\partial
Z_2}\sigma^2_{NZ} + 2 D_{NN}\;,
\end{align}
\begin{align}\label{eq18}
\frac{\partial}{\partial t } {\sigma}^2_{ZZ} = 2 \frac{\partial
\nu_{p}}{\partial Z_2} \sigma^2_{ZZ} + 2 \frac{\partial \nu_{p}}{\partial
N_2}\sigma^2_{NZ} + 2 D_{ZZ}\;,
\end{align}
and
\begin{align}\label{eq19}
\frac{\partial}{\partial t } {\sigma}^2_{NZ} =  \frac{\partial \nu_{p}}{\partial
N_1} \sigma^2_{NN} +  \frac{\partial \nu_{n}}{\partial Z_1}\sigma^2_{ZZ} +
\sigma^2_{NZ}\left(\frac{\partial \nu_p}{\partial Z_2} +\frac{\partial
\nu_n}{\partial N_2}\right).
\end{align}

\begin{figure}[!htb]
\includegraphics*[width=8.6cm]{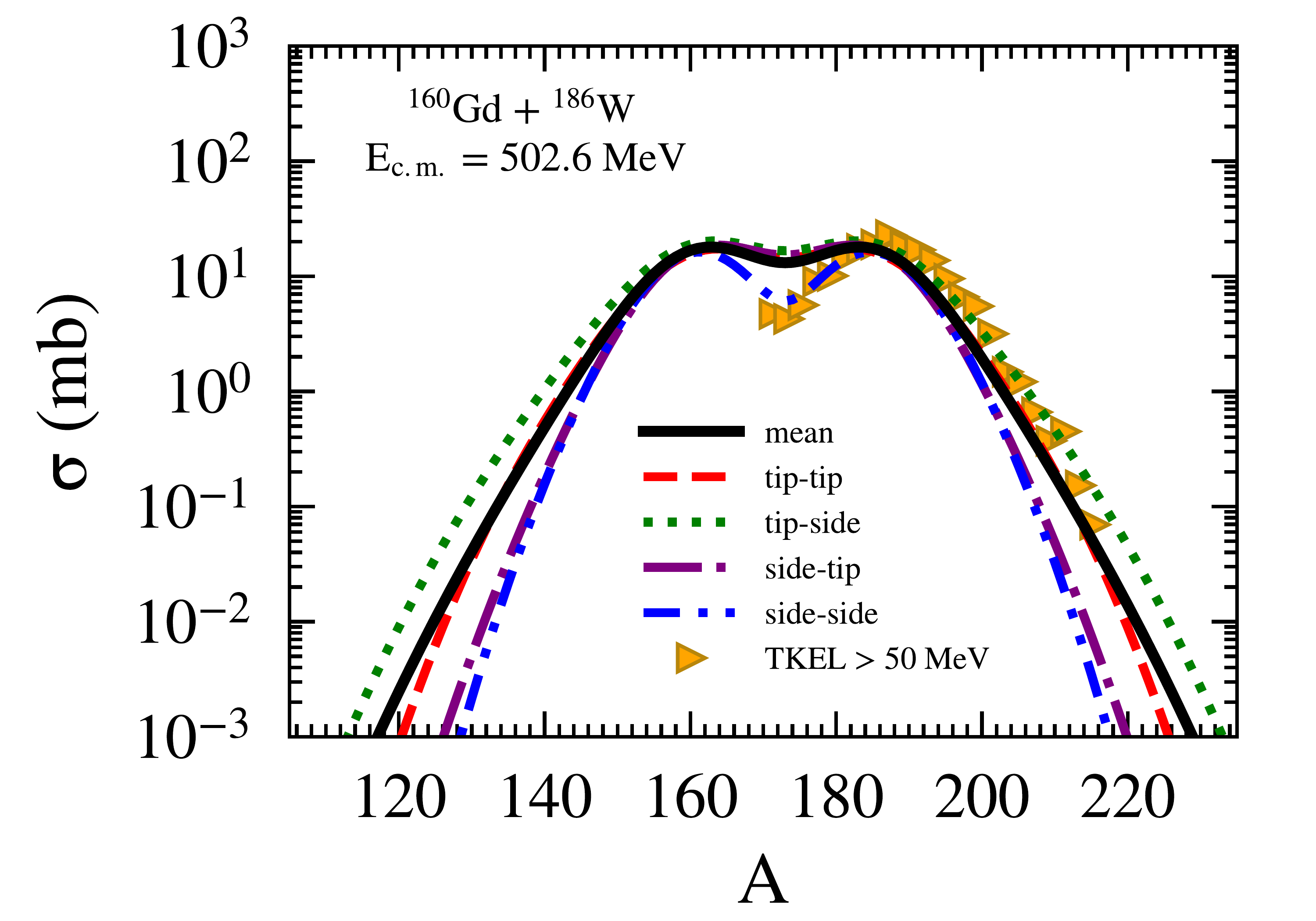}
\caption{Total cross-section in collision of ${}^{160} \text{Gd}+{}^{186} \text{W}$ system at $E_\text{c.m.}=502.6$~MeV
as a function of mass number A. Different geometries are indicated by dashed (red), dotted (green),
dashed-dotted (brown) and dashed-dot-dotted (blue) lines. Average cross-section and experimental data taken from
Ref.~\cite{kozulin2017} are indicated by solid black line and yellow triangles respectively.}
\label{fig5}

\end{figure}
The set of coupled equations are also familiar from the phenomenological nucleon
exchange model, and they were derived from the Fokker-Planck equation for the
fragment neutron and proton distributions in the deep-inelastic heavy-ion
collisions ~\cite{risken1996, schroder1981}. Variances and co-variances are
determined from the solutions of these coupled differential equations with
initial conditions $\sigma^2_{NN}(t=0)=0$, $\sigma^2_{ZZ}(t=0)=0$ and
$\sigma^2_{NZ}(t=0)=0$ for each angular momentum $\ell$. As an example, Fig.~\ref{fig4}
shows neutron, proton dispersions, and mix dispersions in collision of
${}^{160}\text{Gd}+{}^{186}\text{W}$ with initial angular momentum $\ell
=0\hbar$ at $E_\text{c.m.}=502.6$~MeV for different geometries. Probability
distribution of mass number of produced fragments is determined by summing over
N or Z and keeping the total mass number constant $A=N+Z$.

\begin{figure}[!htb]
\includegraphics*[width=8.5cm]{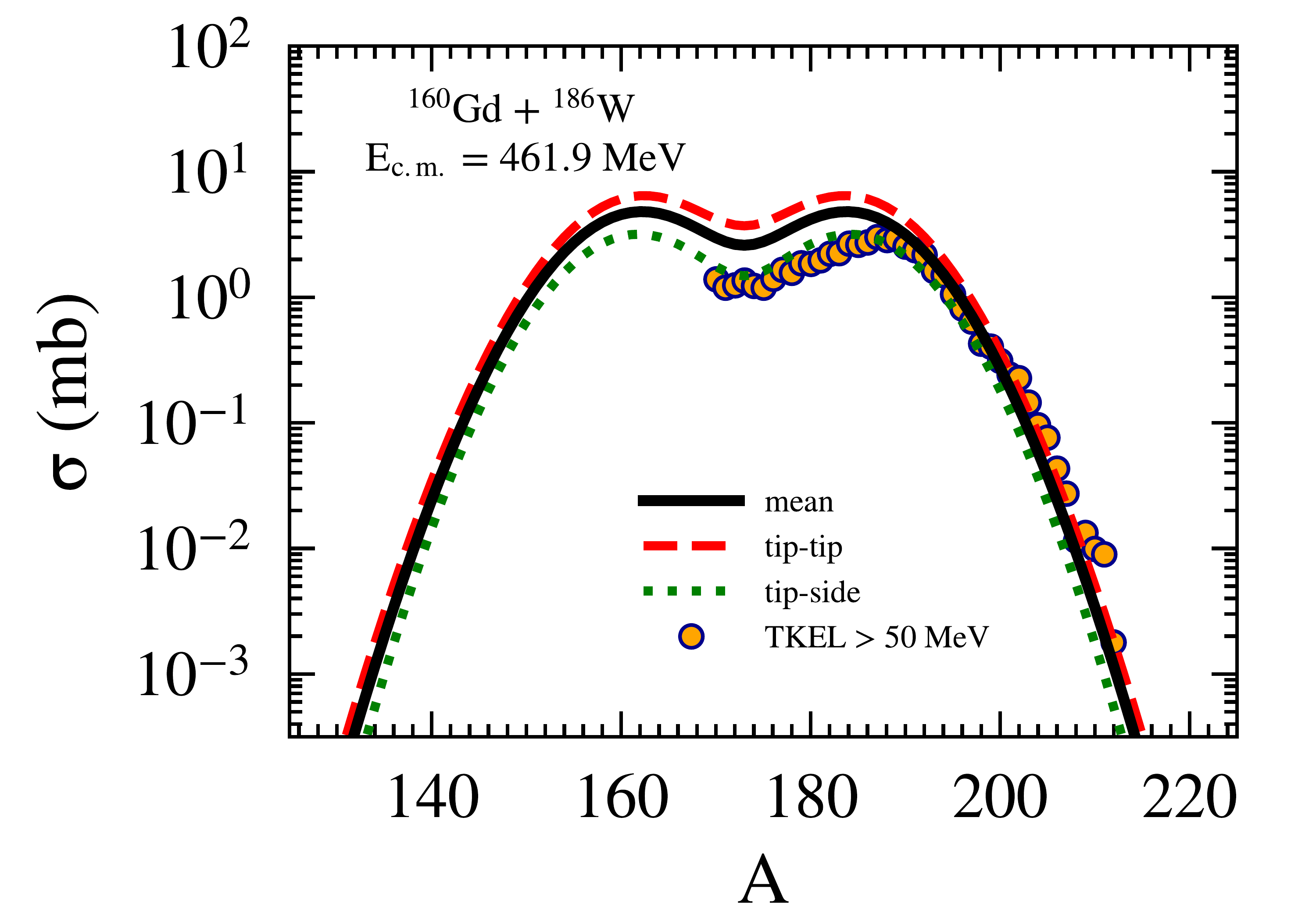}
\caption{Total cross-section in collision of ${}^{160} \text{Gd}+{}^{186} \text{W}$ system at $E_\text{c.m.}=461.9$~MeV  as a function of mass number A. Different geometries are indicated by dotted (green), dashed (red) lines.  Average cross-section and experimental data taken from Ref.~\cite{kozulin2017} are indicated by solid black line and yellow circles respectively.}
\label{fig6}
\end{figure}

\begin{align}\label{eq20}
P_{\ell}(A)=\frac{1}{\sqrt{2\pi}\sigma_{AA}} \exp\left[\frac{-1}{2}\left(\frac{A-A_\ell}{\sigma_{AA}(\ell)}\right)^2\right]\ ,
\end{align}
where mass variance is given by $\sigma_{AA}^2 = \sigma_{NN}^2 + \sigma_{ZZ}^2 + 2\sigma_{NZ}^2 . $
\section{Production cross-sections of primary fragments}
\label{sec4}

It is possible to calculate double cross-sections as a function of neutron and
proton numbers of primary fragments. In experimental analysis of Kozulin
\textit{et al.}~\cite{kozulin2017}, only mass distribution of primary fragments is
published. Therefore,  we present calculation of cross-sections $\sigma^s(A)$ as
function of mass number A of primary fragments using the familiar expression,

\begin{align}\label{eq21}
\sigma ^{s}(A)=\frac{\pi \hbar ^2}{2{\mu}E_\text{c.m.}}\sum
\limits_{\ell_{min}}^{\ell_{max}}(2\ell +1 )  P^{s}_{\ell}(A)\,
\end{align}
with
\begin{align}\label{eq22}
P_{\ell}^{s}(A)=\frac{1}{2}[P_{\ell ,pro}^{s}(A)+P_{\ell,tar}^{s}(A)]\;.
\end{align}
Here label “s” indicates the different geometries of collision. In this
expression, $P_{\ell ,pro}^{s}(A)$ and $P_{\ell ,tar}^{s}(A)$ denotes the
normalized probability of producing projectile-like and target-like fragments.
These probabilities are given by Eq.~(\ref{eq20}) with mean values are
projectile-like and target-like fragments, respectively.
To make the total primary fragment distribution normalized to unity we
multiply by a factor $1/2$.
In summation over initial angular momentum $\ell$, the range of initial orbital angular
momentum depends on the geometry of the detectors in the laboratory frame. In
calculations, we carry out summation over the range from $\ell_\text{min}$ to
$\ell_\text{max}$. The range of $\ell$-values are specified by the angular
acceptance of the detector.  In laboratory frame, the range of acceptance of
detector is $25^\circ - 65^\circ$. The range of $\ell$-values for different
geometries are indicated in Table~\ref{tab1} at $E_\text{c.m.}=502.6$~MeV and
Table ~\ref{tab2} at $E_\text{c.m.}=461.9$~MeV. For $E_\text{c.m.}=461.9$~MeV
energy, there are no $\ell$-value in side-tip and side-side geometries leading
to acceptance range of detector. Fig.~\ref{fig5} shows cross-sections at
$E_\text{c.m.}=502.6$~MeV as a function of mass number A for production of
primary fragments in tip-tip, tip-side, side-tip and side-side geometries with
different colors. Average values of the cross-sections are determined by the
arithmetic mean values $\overline{\sigma}(A) = \frac{1}{4} \sum \limits_{s}
\sigma^s(A)$ of cross-sections at different geometries.  Fig.~\ref{fig6} shows
total cross-sections at $E_\text{c.m.}=461.9$~MeV energy as a function of mass
number A for production of primary fragments in tip-tip and tip-side geometries
with different colors.  Cross-sections in side-tip, and side-side geometries are
nearly zero, and not indicated in the figure. Total cross-section is essentially
determined by tip-tip and tip-side contributions. In this case,  average values
of the cross-sections are determined by the arithmetic mean values of tip-tip
and tip-side contributions, $\overline{\sigma}(A) = \frac{1}{2} \sum \limits_{s}
\sigma^s(A)$ Calculations provide good descriptions of measurements that are
indicated by yellow triangles at high energy data and by yellow circles at
low energy data.

\begin{figure}[!htb]
\includegraphics*[width=8.5cm]{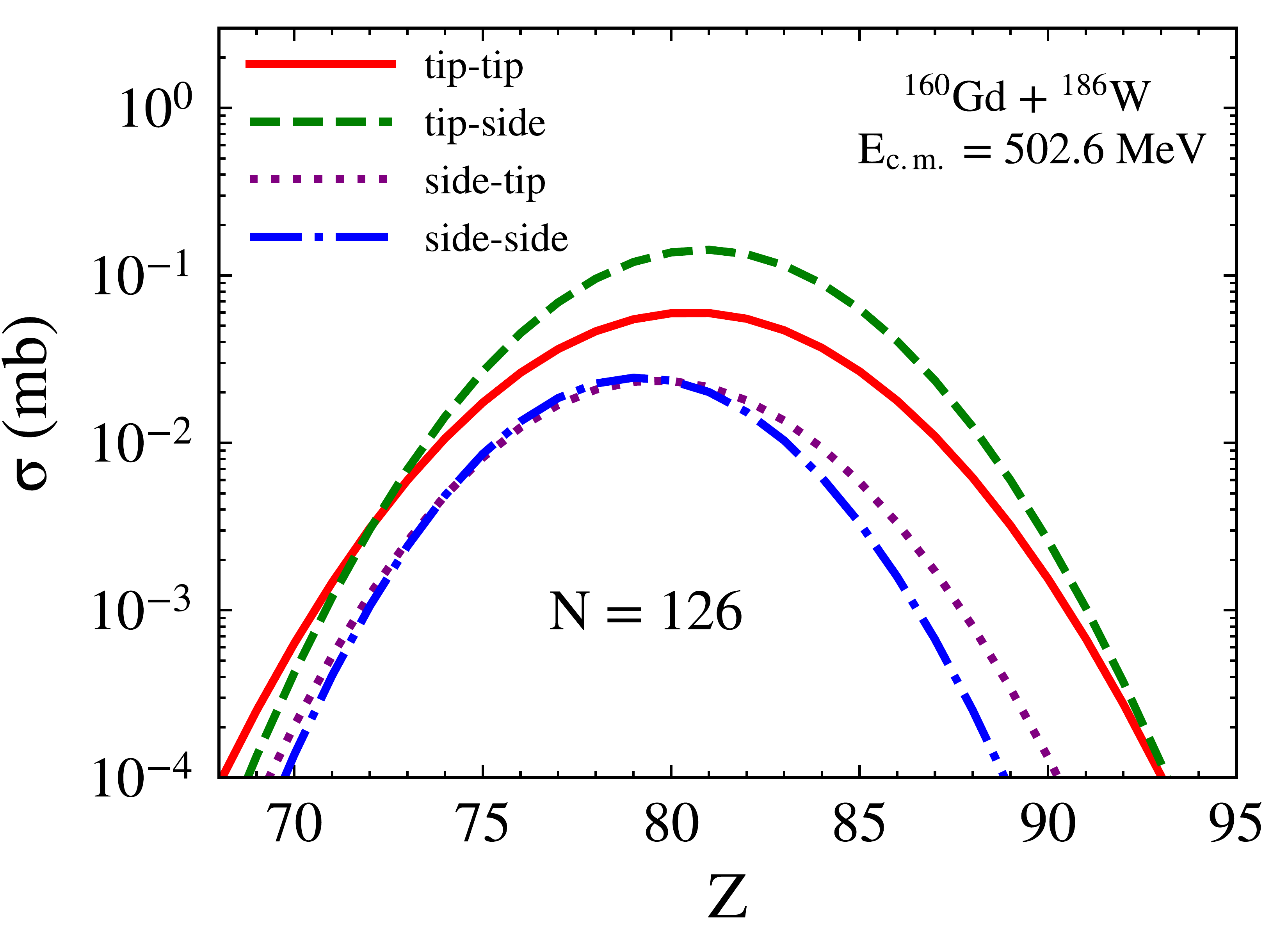}
\caption{Primary product cross sections for $N=126$ isotones in collision of ${}^{160} \text{Gd}+{}^{186} \text{W}$ system at $E_\text{c.m.}=502.6$~MeV  are shown as a function of the proton number of the reaction products. Different geometries are indicated by solid (red), dotted (green), dotted (purple) and dashed-dotted (blue) lines.}
\label{fig7}
\end{figure}

Finally, the prediction of primary production cross sections for $N = 126$
isotones in the ${}^{160} \text{Gd}+{}^{186} \text{W}$ system at
$E_\text{c.m.}=502.6$~MeV are given in Fig.~\ref{fig7}. It is observed that the
primary product cross sections in the tip-side geometry are roughly one order of
magnitude higher than those in the side-tip and side-side geometries,
highlighting the effect of inverse quasi-fission observed in the tip-side
geometry. Results clearly demonstrate that depending on the deformation and
relative orientation of the reaction partners, outcome of the reaction differs.
This effect supports the idea that MNT reactions can serve as a useful tool to
produce neutron rich heavy nuclei near the lead region, which may not be
available via fusion-fission and fragmentation.

\section{Conclusions}
\label{sec5}

We investigate multi-nucleon transfer mechanism in quasi-fission reactions of
${}^{160}\text{Gd}+{}^{186}\text{W}$ system at $E_\text{c.m.}=502.6$~MeV and
$E_\text{c.m.}=461.9$~MeV employing quantal diffusion description based on the
Stochastic Mean-Field approach.  We evaluate transport coefficients associated
with charge and mass asymmetry variables in terms of time-dependent
single-particle wave functions of TDHF theory. Transport description includes
quantal effect due to shell structure, full geometry of the collision dynamics
and Pauli Exclusion Principle and does not involve any adjustable parameters
aside from the standard description of the effective Hamiltonian of TDHF theory.
In transport description, in addition to diffusion coefficient, we need to
determine first and second derivatives of the potential energy surface with
respect to neutron and proton numbers. It is possible to determine iso-scalar
and iso-vector curvature parameters in terms neutron and proton drift
coefficients and diffusion coefficients with the help of Einstein relations. Joint
probability distribution of primary fragments is determined by a correlated
Gaussian function in terms of mean values of neutron-proton numbers and neutron,
proton, mixed dispersions for each initial angular momentum. We calculate
cross-sections of primary fragments as a function of mass number and compare
with data of Kozulin \textit{et al.}~\cite{kozulin2017}. Calculations provide good
description of primary mass distributions at both bombarding energies.

\begin{acknowledgments}
S.A. gratefully acknowledges Middle East Technical University for warm
hospitality extended to him during his visits. S.A. also gratefully acknowledges
F. Ayik for continuous support and encouragement. This work is supported in part
by US DOE Grant Nos. DE-SC0015513 and DE-SC0013847.
This work is supported in part by TUBITAK Grant No. 122F150. The numerical
calculations reported in this paper were partially performed at TUBITAK ULAKBIM,
High Performance and Grid Computing Center (TRUBA resources).
\end{acknowledgments}
\clearpage
\begin{widetext}
\appendix*

\renewcommand{\thefigure}{\arabic{figure}A}
\setcounter{figure}{0}
\section{}
\label{appA}

We can estimate the averaged values of reduced isoscalar and isovector
curvature parameters with the help of Einstein relations, Eq.~(\ref{eq3}), in the overdamped limit. We evaluate the average values of the reduced
curvature parameters for different geometries by carrying out the time integrals in
Eqs.~(\ref{eq5}-\ref{eq10}) over suitable time intervals. These time
intervals are indicated in following figures for
$^{160}\text{Gd}+^{186}\text{W}$ and $^{172}\text{Gd}+^{174}\text{W}$
reactions at $E_\text{c.m.}=502.6$~MeV.
\begin{figure*}[!htb]
\includegraphics*[width=12cm]{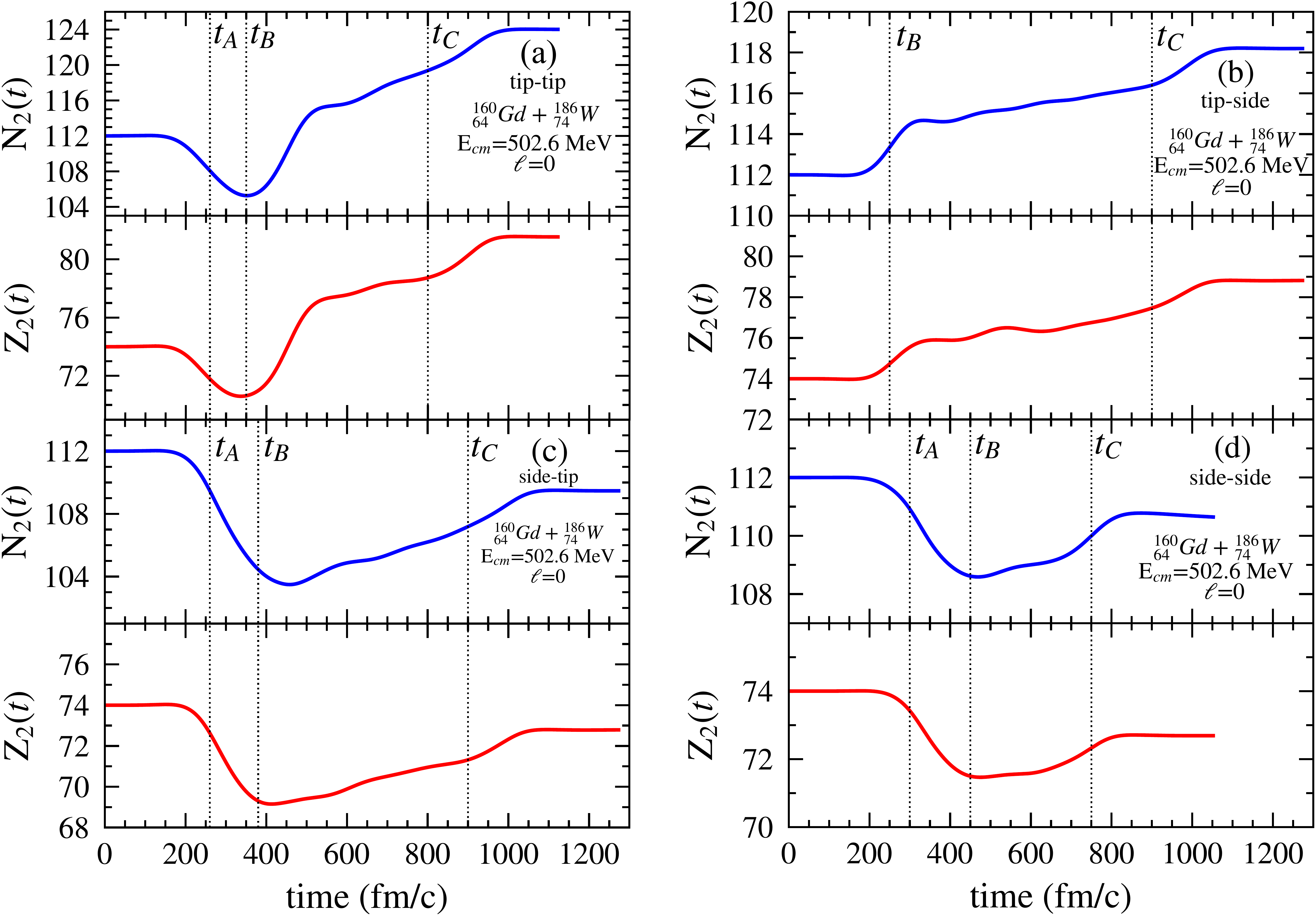}
\caption{Mean-values of neutron and proton numbers of W-like fragments in the
head-on collision of ${}^{160}\mathrm{Gd}+{}^{186}\mathrm{W}$ system at $E_\mathrm{c.m.}=502.6$~MeV in tip-tip
(a), tip-side (b), side-tip (c) and side-side (d) geometries.}
\label{fig1A}
\end{figure*}

\begin{figure*}[!htb]
\includegraphics*[width=12cm]{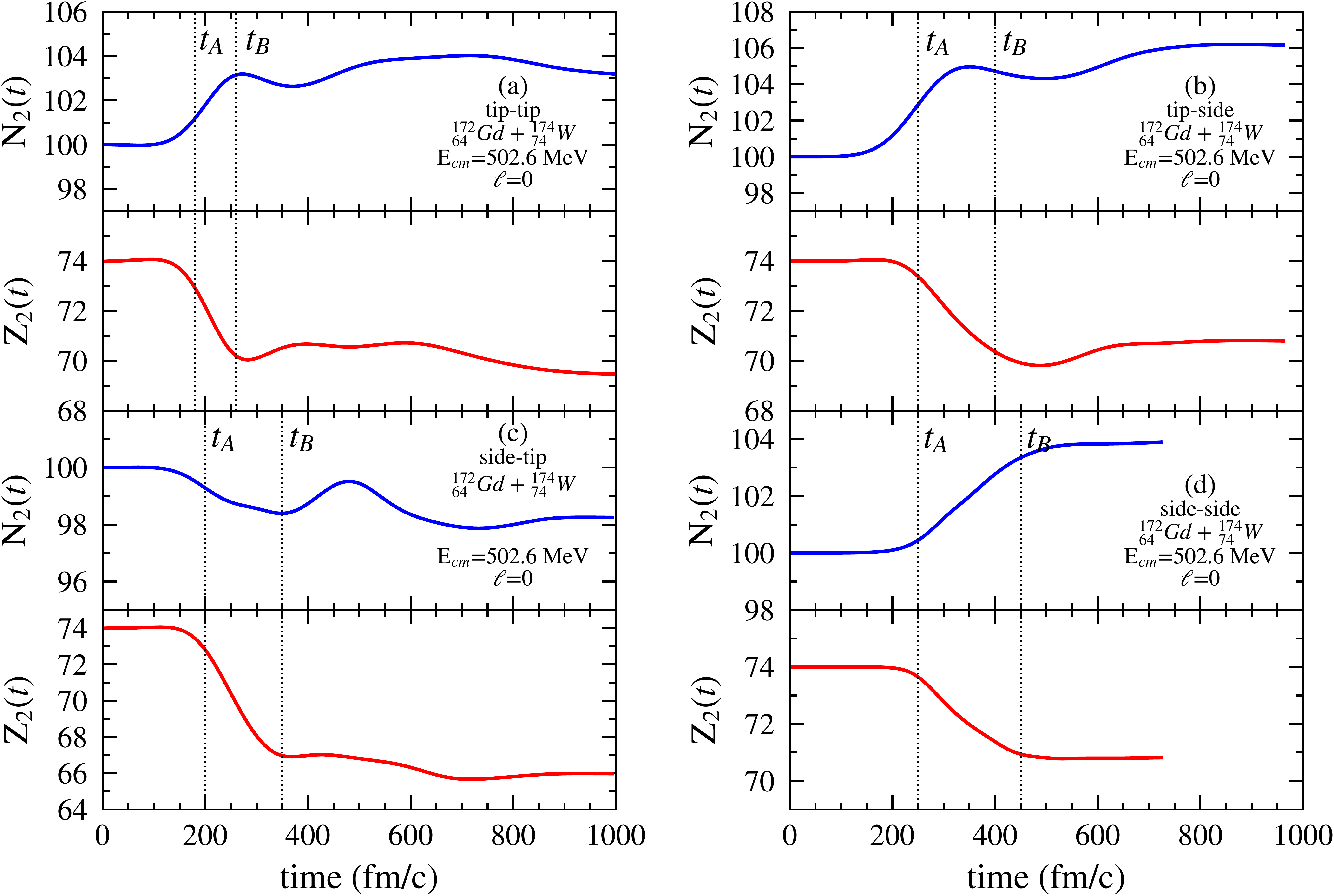}
\caption{Mean-values of neutron and proton numbers of W-like fragments in the
head-on collision of ${}^{172}\text{Gd}+{}^{174}\text{W}$ system at $E_\text{c.m.}=502.6$~MeV  in tip-tip
(a), tip-side (b), side-tip (c) and side-side (d) geometries.  }
\label{fig2A}
\end{figure*}

\begin{figure*}[!t]
\includegraphics*[width=12cm]{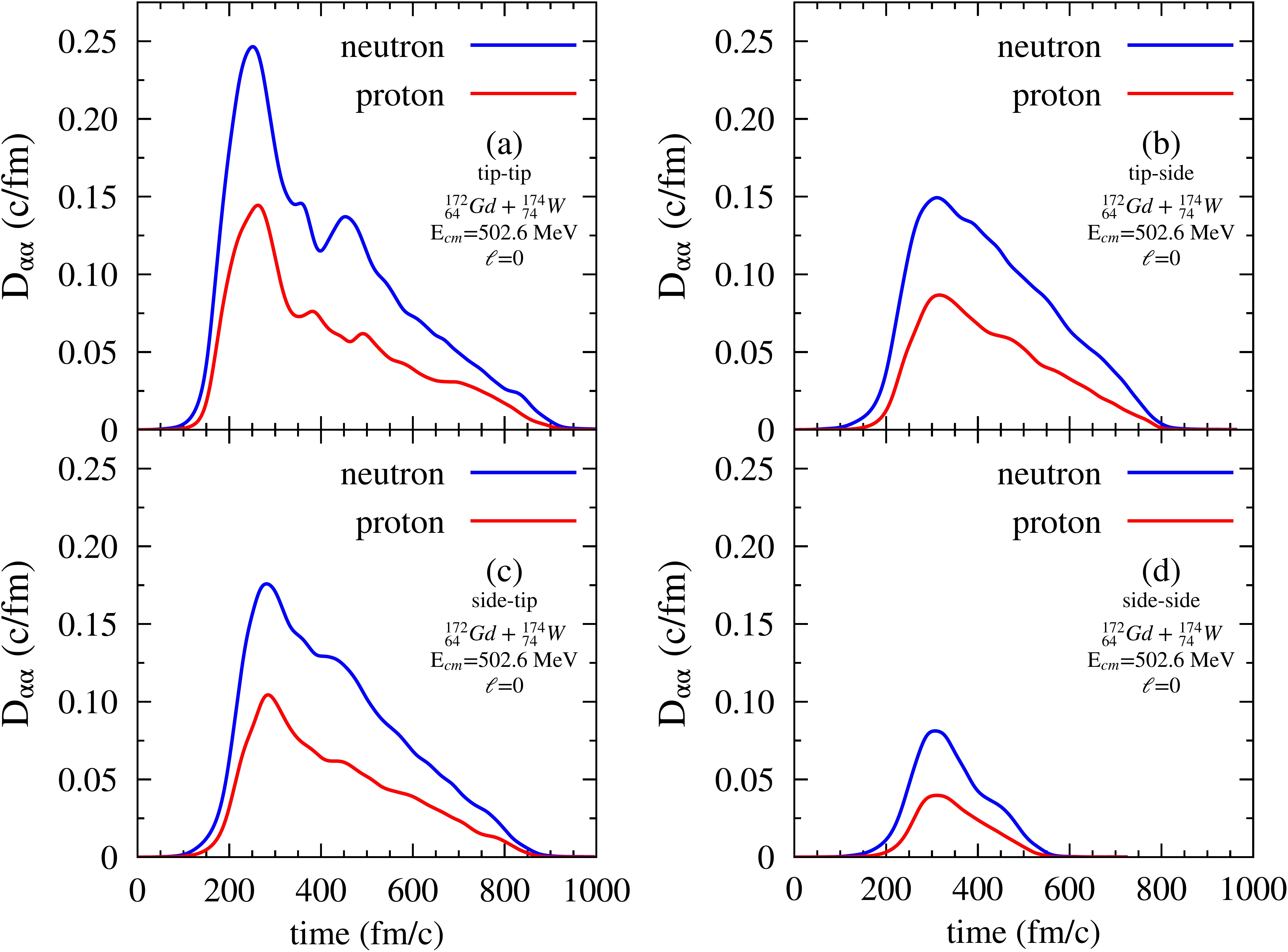}
\caption{Diffusion coefficients in the head-on collision of
${}^{172}\text{Gd}+{}^{174}\text{W}$ system at $E_\text{c.m.}=502.6$~MeV  in tip-tip
(a), tip-side (b), side-tip (c) and side-side (d) geometries. }
\label{fig3A}
\end{figure*}

\end{widetext}
\clearpage
\bibliography{VU_bibtex_master}

\end{document}